# Resolving buried interfaces with Low Energy Ion Scattering

Adele Valpreda[1,a)], Jacobus M. Sturm[1], Andrey Yakshin[1], Marcelo Ackermann[1]

[1]Industrial Focus Group XUV Optics, MESA+ Institute for Nanotechnology, University of Twente, Enschede, the Netherlands

a) a.valpreda@utwente.nl

We investigate the use of Low Energy Ion Scattering (LEIS) to characterize buried interfaces of ultra-thin films. LEIS spectra contain depth-resolved information in the so-called sub-surface signal. However, the exact correlation between the sub-surface signal and the sample's depth composition is still unknown. For this reason, LEIS spectra so far only provided qualitative information about buried interfaces.

In this study, we investigate nm-thin films of Si-on-W and Si-on-Mo, where we compare simulated data to LEIS spectra. We present a method to extract depth-sensitive compositional changes – resolving buried interfaces - from LEIS spectra for the first few nanometers of a thin film sample.

In the case of Si-on-Mo, the simulation of the LEIS sub-surface signal allows obtaining a quantitative measurement of the interface profile that matches the value determined using the LEIS layer growth profile method with an accuracy of 0.1 nm. These results pave the way to further extend the use of LEIS for the characterization of features buried inside the first few nanometers of a sample.

## I.  INTRODUCTION

Ultrathin films of only a few nm pose unique challenges in the characterization of interfaces. When film thicknesses are a few nanometers at most, the interface makes up a major part of the final structure, and hence determines many of the film's properties. To unravel, and ultimately predict the properties of such thin films, characterizing the interface composition with quasi-atomic accuracy is key.

Several methods can be used to probe the interface quality but no method is free of issues. Commonly used methods include transmission electron microscopy (TEM) which typically requires extensive experimental effort and X-ray photoelectron spectroscopy (XPS) which offers a limited depth resolution due to large information depth. Low Energy Ion Scattering (LEIS), XPS and secondary ion mass spectrometry (SIMS) can also be used in combination with sputter depth profiling, which, however, will introduce sputtering artifacts.

In this paper, we present the use of the LEIS sub-surface signal for the characterization of buried interfaces in a static mode. This is interesting because it avoids the use of sputtering steps, which are currently the limiting factor for the use of LEIS to resolve buried features with quasi-atomic resolution.

Along with quantification of the composition of the outermost atomic layer, LEIS provides compositional information about deeper layers, down to ca. 10 nm. These two signals are distinguishable in LEIS measurements as a peak-like 'surface' signal and a background 'sub-surface' signal respectively, as it can be seen in figure 3. The presence and intensity of the sub-surface signal depend on the chemistry of the surface and target and projectile conditions such as the mass of the target atoms and the mass and energy of the projectiles [1-4].

The surface selectivity of the peaks in LEIS spectra enables the characterization of the change in surface coverage as a function of the as-deposited film thickness, the so-called LEIS layer growth profile. The procedure used to record the LEIS layer growth profiles is described in [5]. In the studies [5-11], the authors made use of LEIS layer growth profiles to characterize the nanolayer structure evolution and intermixing behavior of Transition-metal/silicon (TM/Si) thin-film structures, Transition-metal/Transition-metal (TM/TM) structures and Transition-metal oxides deposited by magnetron sputtering and Atomic Layer Deposition. In the studies [5, 7, 9] the authors showed the effectiveness of the error function and the logistic function to describe the interface profile in thin films. In the study [9], the layer growth profile of a comprehensive set of TM/TM structures allowed the authors to derive empirical rules to qualitatively predict the growth characteristics of the system based on atomic size difference, surface-energy difference, and enthalpy of mixing between the film and substrate atoms.

In LEIS layer growth profiles, the fact that the interface is characterized while being formed limits the use of the method to systems that are not subjected to matrix effects and segregation. Specifically, segregation during growth results in a mismatch between the as-deposited surface composition and the final interface profile. For these reasons, in recent years the sub-surface signal has gained more attention with the aim of improving the static non-destructive depth analysis of sample compositions, the so-called LEIS static depth profiling, offering an alternative to the layer growth profile in modern thin film science.

It was shown that it is possible to determine the thickness of a top film with sub-nm resolution from the shape of the sub-surface signal in LEIS measurements, with the restriction that the difference in mass between the top film and substrate needs to be sufficiently large to separate their respective contributions [2-4, 8, 12-17]. The method was successfully demonstrated for the combination of $ZrO_2$ and Si [17].

In literature, several authors have already shown that Monte Carlo calculations performed with the TRBS code [18] can provide valid simulation of LEIS data [3, 14, 19-25]. The study by Brüner et al. [14] specifically showed that TRBS simulations are a valuable tool for film thickness analysis. However, the authors state that for the investigated structures, allowing for layer intermixing in TRBS does not significantly change the outcome of the simulation. From these results, it seemed impossible to measure an interface width by LEIS spectrum analysis paired with TRBS simulations.

In LEIS measurements, the projectiles' energy loss due to the interaction with the electrons is stochastic and therefore subjected to depth-dependent straggling. Although it is true that TRBS offers the possibility to include electronic straggling in the simulation, one must consider that when we apply TRBS to the LEIS regime (of a few keV) the electronic straggling is overestimated by the code, which is tailored to the MeV regime [14]. For this reason, past attempts to simulate LEIS data from TRBS calculations either included a custom-made model of electronic straggling or manually adjusted the TRBS smoothing function.

To the authors' knowledge, the models used so far for the simulation of electronic straggling did not take into account the dependence of electronic straggling on energy. The risk with this simplification is to overestimate the electronic straggling in the high energy side of the spectrum (which correspond to lower penetration depth). The implementation of an overestimated smoothening function can explain why the simulations appear insensitive to the small compositional changes that are present below the surface of the sample.

In this study, we measure the error in the simulation of LEIS spectra when no electronic straggling is applied, aiming to improve the understanding of electronic straggling in the LEIS regime. We then explore the characterization of a buried interface by comparing the experimental and simulated LEIS sub-surface spectra.

We use W/Si and Mo/Si thin films as model structures. W/Si structures are expected to have a relatively sharp and stable interface when Si is deposited on W [11]. As such, they are a good example structure for assessing the contribution of electronic straggling to the shape of the sub-surface signal in LEIS spectra. The results show that the electronic straggling is a function of the penetration depth of the ions inside the sample.

Mo/Si thin-film structures are expected to have a relatively broad interface when Si is deposited on Mo [10], which makes them a good model structure for assessing the contribution of interface width to the shape of the sub-surface signal. We show that the method of comparing the experimental and simulated LEIS spectra is sensitive to the interface width in the case of short penetration depths, where the effect of electronic straggling is reduced to the minimum.

## II. EXPERIMENT

### *A. Deposition*

All samples were fabricated in a home-designed ultra-high vacuum (UHV) system (base pressure $<1\times10^{-9}$ mbar) which allows in-vacuum transfer between the thin film deposition chamber and the LEIS analysis chamber.

The following structures were deposited, a 30 nm silicon film for the measurement of silicon reionization function, three Si-on-W structures and one Si-on-Mo structure for the characterization of buried interfaces. All the structures were deposited onto super-polished Si substrates with native oxide. The bi-layer structures for interface characterization are shown in figure 1.

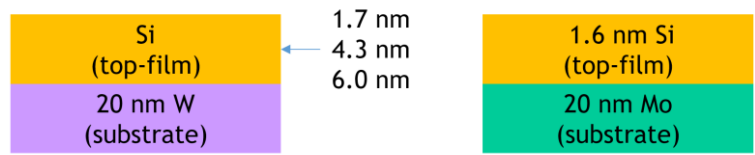

*FIG. 1. Bilayer structures used for LEIS characterization of buried interfaces. Three structures were used with Si-on-W. The thickness of the Si top film varied between the structures, while the deposition parameters were kept constant. One structure was used with Si-on-Mo.*

All the films were deposited at room temperature using magnetron sputtering. The argon process gas working pressure was $0.6\times10^{-3}$ mbar. The substrate-to-target distance was 8 cm for all materials. To prevent cross-contamination, all magnetrons were equipped with a shutter.

W and Mo were deposited by direct current (DC) magnetron sputtering. The sputter powers used were 12 W and 10 W respectively. The corresponding sputter voltages were 357 V and 338 V, and the deposition rates were 0.07 nm/s and 0.11 nm/s.

The settings used for silicon varied between samples. Note that for the silicon films of interest for this study, the surface roughness is not expected to vary depending on the deposition settings. For the ion-fraction and Si-on-W samples, DC sputtering was used. The sputter power value was 12 W, matching the settings used in the study by Zameshin et al. [11]. The corresponding Si sputter voltage was 437 V, and the Si deposition rate was 0.05 nm/s. For the Si-on-Mo sample, radiofrequency (RF) sputtering was used for the deposition. The sputter power was 30 W, matching the settings used in the study by Reinink et al. [10]. The corresponding Si deposition rate was 0.02 nm/s.

To monitor the deposited thickness, all magnetrons are equipped with a quartz crystal microbalance (QCM) which is calibrated against ex-situ X-ray reflectivity (XRR) measurements of reference layers. Note that magnetron sputtering produces films that are very close to bulk density (in the 98-99% range), therefore the deposited mass can be related to thickness.

For Si-on-W, the thickness of the top film was determined using two methods: 1) the QCM, and 2) LEIS static depth profiling. The agreement between the two measurements is ±0.3 nm. For Si-on-Mo, the thickness of the top film was measured by LEIS static depth profiling only.

### *B. LEIS Characterization*

LEIS measurements were performed using an IONTOF GmbH Qtac100 high-sensitivity LEIS spectrometer with a base pressure of $1 \times 10^{-10}$ mbar.

The system is equipped with two electron-impact ion sources (primary source and sputter gun), a double toroidal electrostatic analyzer (DTA), and a position-sensitive detector. The primary source and sputter gun are positioned at incidence angles of 0° and 59° with respect to the sample surface normal. The DTA detects ions that are backscattered at an angle of 145°.

During the measurement, the primary beam rasters over a 1x1 mm$^2$ area. For the study of the silicon reionization function, a 6 keV He$^+$ beam with a 4 nA current was used for the first measurement. He$^+$ beams of 5 keV, 4 keV, and 3 keV were also used. The measured beam currents were 4.3 nA, 4.2 nA and 3.1 nA, respectively. For all the measurements, the acquisition time was under 4 min with an ion dose of $2 \times 10^{15}$ ions/cm$^2$. For the interface characterization, a 3 keV He$^+$ beam with a 3 nA average current was used for measurements. The acquisition time was around 3 min with an ion dose of around $3.5 \times 10^{14}$ ions/cm$^2$.

Whenever sputtering was performed, a 0.5 keV Ar$^+$ beam with a 100 nA average current was used over a raster area of $2 \times 2$ mm$^2$.

## III. TRBS SIMULATIONS

For this study, we used the Monte Carlo code TRBS which is a specialized version of the TRIM code [26], optimized for the calculation of backscattered particles [18]. We used the version of TRBS implemented into the IONTOF SurfaceLab software (I-TRBS).

### *A. Working principle*

The code models the trajectory of ions inside a target as formed by free paths between nuclei and scattering events with the nuclei.

In a free path, the partial energy loss resulting from the interaction with the target's electrons (electronic stopping) is implemented. Previous studies [14, 19-21] showed that electronic stopping is typically underestimated by TRBS when performing simulations with low-energy ions. To compensate for this, TRBS requires the user to specify a correction for the electronic stopping (ESC values).

In a scattering event, the universal scattering potential is used to model the scattering probability. To mimic the experimental condition while limiting the computational costs, TRBS solves individual scattering integrals only when the scattering angle is above a user-defined cutoff angle. The collisions resulting in smaller scattering angles are accounted for globally as a continuous nuclear energy loss [18].

Biersack et al. provide a detailed description of TRBS [18], including the calculation methods used by the algorithm. Brüner et al. [14] provide a detailed description of the adjustments to make in order to use the TRBS code to simulate LEIS spectra.

The input for the program is a file where the ion species, the primary energy of the ions, and the target composition are defined. For each layer of the target, the user specifies the stoichiometry, thickness, density, ESC, and screening length correction (SLC). The latter is a correction factor for the empirical scattering potential which affects the intensity of the simulated spectrum. The parameters ESC, SLC, and cutoff angle are further discussed in the section Calculation Details.

As output, TRBS gives two energy spectra of backscattered particles. The first is the particles' energy distribution without the influence of electronic straggling (uncorrected spectrum). The second is the result of the uncorrected spectrum convoluted with a Gaussian energy distribution where the standard deviation represents the mean electronic straggling width for each channel (corrected spectrum) [18]. Previous studies showed that the corrected spectrum often overestimates the influence of electronic straggling when the primary energy is of the order of keV [14, 18]. For this reason, in the study by Brüner et al. [14] the straggling correction is custom-made with a gaussian that has a Full With at Half Maximum (FWHM) of around 300 eV. The result is a good fit of the spectra but the applied uniform broadening leads to simulations that are insensitive to the interdiffusion between thin films. To observe the effect of electronic straggling on the LEIS spectra, we compare the uncorrected spectra with LEIS measurements, as shown in section VI.

The main difference between LEIS experiments and TRBS simulations is that the projectile charge state is not included in TRBS simulations. This is why TRBS simulations of scattered particles have no contrast between surface and sub-surface signals. This difference is the key feature for the calculation of the reionization function of a material by means of TRBS simulations, as described in section IV.

### *B. Calculation Details*

For each simulation, the ion species, the energy of the ions, and the target composition were chosen to match a corresponding LEIS experiment.

For each material used in this study, the value of the ESC parameter was measured by performing LEIS measurements and TRBS simulations on thin films of known thickness. For silicon, the measurements were performed at three different primary energies, 3 keV, 4 keV, and 5 keV, and on three samples of different thickness for better accuracy. The Si-ESC factor obtained is valid for all the investigated primary energies and thicknesses. For W and Mo, only one measurement with 3 keV primary energy was performed for the evaluation of the ESC. Note that Si-ESC is critical for the measurement of Si reionization function, while W and Mo ESCs will only affect the low energy side of the spectra used in this study, which is of no interest for the measurement of the interface width. The exact values reported in Table I were used for the ESC factors in all the simulations presented in this study.

TABLE I. Electronic stopping correction (ESC) for TRBS simulations of the materials used in this study. The measurements were performed on films of known thickness.

| Material | ESC (dimensionless) |
|---|---|
| Si | 2.2 ±0.1 |
| W | 1.9 ±0.3 |
| Mo | 1.9 ±0.3 |

The parameter SLC did not significantly change the shape of the spectra studied in this work. Therefore, we used the default value for low energy equal to 0.85. For the cutoff angle, the default value of 0.08 was assigned to the corresponding simulation parameter during a first investigation step. This allowed to obtain fast results with a typical computation time below 500 s. In a second step, simulations were run with a much lower cutoff angle. This led to identical results with the only difference of reduced noise. A total number of $10^8$ ions were used in each simulation, this was sufficient to achieve smooth simulation results.

### C. Adjustments to the TRBS spectra

Instrumental broadening is a contribution to the shape of LEIS spectra that is not simulated by TRBS. We implemented a simple approximation of instrumental broadening through a Gaussian convolution of the output spectrum. The width of the Gaussian was taken equal to the width of the surface peak i.e. around 50 eV, assuming that the width of the surface peak represents the minimal broadening that is also expected for in-depth information.

In TRBS output spectra, in units particles/total particles, the yield depends on the channel size. The wider each channel the more particles are included in it. We hence normalize the result by the channel size. The resulting spectrum of backscattered particles, in units particles/(total particles*eV), has the same maximum yield regardless of the resolution.

TRBS simulations allow isolating the signal coming from the first layers of the sample. We compared such TRBS energy spectra with the corresponding LEIS spectra and noticed that in our case there is a mismatch of around 70 eV. This is attributed to both the inelastic energy loss of the reionization process and the energy calibration of the LEIS experiments. For each structure presented here, we shifted the experimental spectra accordingly, obtaining aligned experimental and simulated spectra.

## IV. MODEL FOR REIONIZATION FUNCTION

From a physical point of view, we make the following assumptions regarding charge transfer between projectiles and target atoms:

- Noble gas ions penetrating the target get neutralized
- Detected ions scattered from the subsurface are reionized at the surface upon leaving the sample
- For a given surface chemistry, the probability for a projectile to be reionized at the surface is a function of the final energy

With these assumptions, the reionization ion-fraction as a function of energy (reionization function) can be calculated by dividing the LEIS spectrum of backscattered ions by a spectrum of backscattered particles. The latter can be calculated either from single scattering approximations or by Monte Carlo calculations such as TRBS simulations. The difference is that Monte Carlo simulations take into account multiple scattering which is a key factor contributing to the shape of LEIS spectra.

The method of calculating the reionization function by dividing the LEIS spectrum by the corresponding TRBS simulation, first presented in 2015 by Brüner et al. [14], was recently used to investigate the ion fraction of oxides [19-21] and is used in this study to obtain the silicon reionization function. The result is shown in figure 2.

To our knowledge, there are no quantitative models describing how the reionization function scales as a function of energy. For this reason, it is difficult to identify in which energy range the signal from sputtered atoms has a significant contribution to the sub-surface signal. A possibility is to perform Time of Flight measurements. However, this was not enough to avoid the detection of sputter atoms in previous studies [14]. For this reason, we calculated the maximum energy of sputtered silicon for each primary energy from elastic kinematics, as described in the Appendix, and excluded the data below such values.

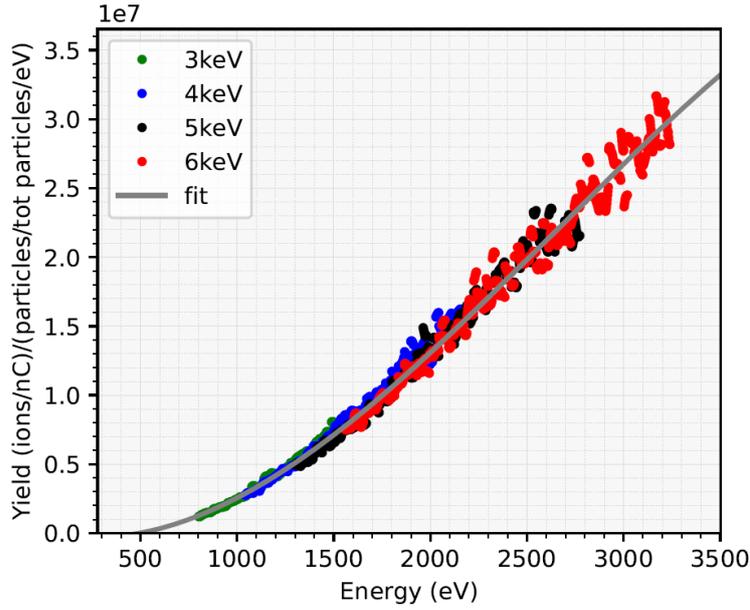

FIG. 2. Silicon reionization function determined with four different primary ion energies. For each primary energy, the reionization function was obtained on a 30 nm silicon film as the point-to-point ratio between the LEIS experiment and TRBS simulation. The data is fitted with a polyline of degree 3.

The reionization probability increases as a function of the final energy of the projectile. This is expected, the higher the final energy, the closer the projectile can get to the target atom during the last collision, and the higher the probability for charge transfer [1, 3, 14, 21]. In addition, higher final energy implies a higher probability that a reionized particle survives Auger neutralization in the path toward the detector.

The reionization functions determined with four different primary energies overlap, showing that the ion fraction does not depend on the primary energy. This is expected considering that the final reionization happens at the surface when the projectile is about to leave the sample.

The reionization energy threshold resulting from the calculation is in agreement with the measurements included in the review by Brongersma et al. which reported a threshold between 300 eV and 500 eV [1].

## V. MODEL FOR LEIS SUB-SURFACE SIGNAL

When focusing on bi-layer structures whose surface is fully closed by atoms of the top film, the reionization ion-fraction of the top material describes the reionization probability of any projectile, including those that were backscattered by atoms of the substrate. This is based on the assumption that the final reionization happens at the surface. Therefore, for a given bilayer structure of known film thicknesses, multiplying the spectrum of backscattered particles (such as the TRBS spectrum) by the reionization function of the top film gives a simulation of the LEIS spectrum with the exception of the surface peaks. The latter are mainly formed by ions surviving neutralization during surface backscattering and cannot be simulated by the reionization function.

Figure 3 shows the steps for simulating the sub-surface signal (also called background) of a LEIS spectrum as implemented in this study. The structure of 1.7 nm Si-on-W was used in this case. The primary energy of the ions was 3 keV.

The resolution of the experiment was lowered to match the resolution of the TRBS simulation which in this case is a channel size of 25 eV. We start from the uncorrected TRBS spectrum. The convolution of the latter by a Gaussian with a 50 eV width allows us to obtain the spectrum of backscattered particles in units particles/(total particles*eV) (figure 3, green). We multiply the spectrum of backscattered particles by the silicon reionization function, thereby obtaining a spectrum of backscattered ions in units ions/nC (figure 3, blue). We multiply this by a scaling factor to take into account the detection efficiency in the experiment and obtain a simulation of the LEIS sub-surface signal (figure 3, red).

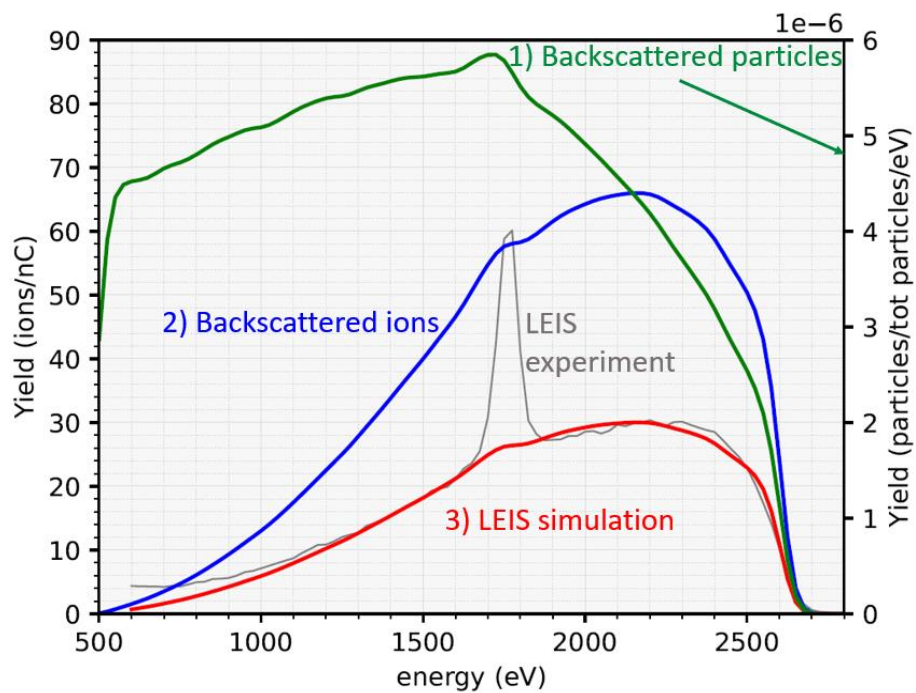

*FIG. 3. Comparison of experimental LEIS spectrum (gray), TRBS spectrum of backscattered particles (green), spectrum of backscattered ions (blue), and simulation of LEIS sub-surface signal (red) for the structure 1.7 nm Si on 20 nm W. The experiment was shifted in energy to match the surface signal of the simulation. The uncorrected TRBS spectrum was normalized by the channel size and convoluted with a 50 eV FWHM Gaussian to simulate instrumental broadening; the TRBS yield should be read on the right side axis. The spectra of backscattered ions and the simulation of LEIS sub-surface signal (background simulation) have the same unit as the LEIS experiment; the corresponding yield should be read on the left side axis. The numbers in the figure indicate the sequence of steps implemented in this study for simulating the sub-surface signal of a LEIS spectrum.*

Below 1200 eV the LEIS signal starts to deviate from the simulation. This is attributed to the contribution of sputtered Si atoms to the ion signal. In the case of Si-on-W we expect the maximum energy of sputtered Si atoms will be higher compared to the case of pure Si (described in the Appendix). This is due to the fact that the projectiles can backscatter on W and then create a Si recoil in a second collision. He projectiles backscattered on in-depth W have about 1.5 times more kinetic energy compared to scattering on Si (equation 2), this will produce higher energy sputtered Si atoms.

Accurately modeling the LEIS sub-surface signal with the method described above is valuable since its shape provides information relevant to depth resolution and surface quantification. In the section 'results and discussion', we further investigate these two interesting features of the LEIS spectra.

# VI. RESULTS AND DISCUSSION

## A. Influence of electronic straggling

In our model, we neglect the effect of electronic straggling on the LEIS spectrum. Therefore, any influence of electronic straggling is found in the residual error between the simulation and measured data.

We compare the simulations of three Si-on-W structures which present different thicknesses of the Si top film. When depositing Si on W, we expect to obtain a relatively sharp and stable interface [11]. Between the three samples, the width of the interface is expected to be constant as the deposition settings were kept constant for the three depositions. The surface roughness is expected to be sufficiently similar between the three samples, considering the amorphous structure of the silicon film. For a thicker film, the effect of electronic stopping and electronic straggling should be higher, therefore we expect to see an increasing error between simulation and data for increasing thickness of Si top-film.

To compare the results, we determine the relative error in the fit of the W signal at high energy, which corresponds to the signal coming from interfacial W. The results for the three structures are shown in figure 4. The corresponding values of the relative error at high energy are reported in table II. The same trend holds if we consider the whole spectrum for the calculation of the relative error.

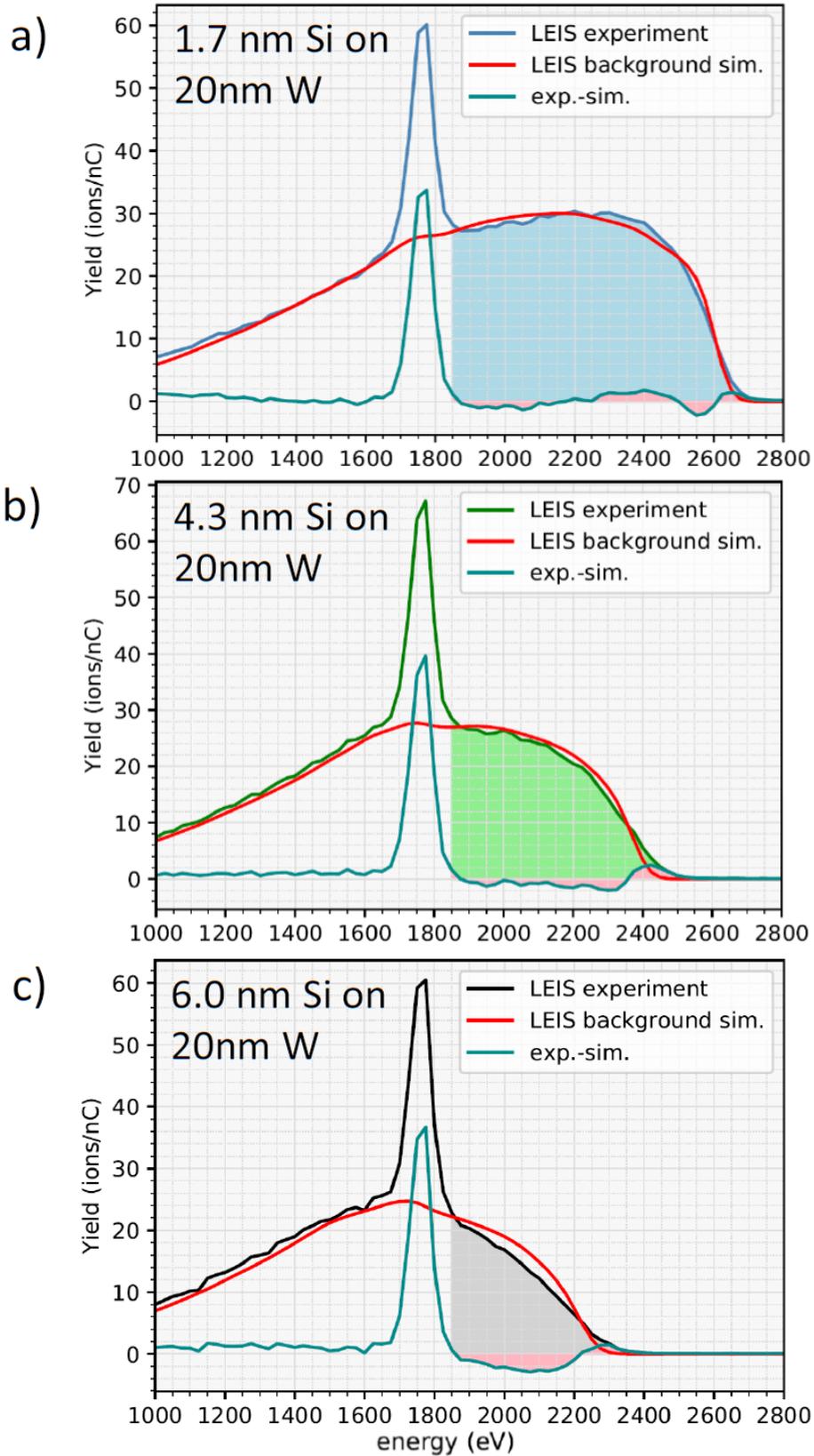

FIG. 4. LEIS experiment with 3keV He+ ions compared to the corresponding simulation for the different structures in a) b) and c). The interface is modeled as infinitely sharp in the simulation. The residual error is calculated as the difference between the experiment and simulation for each point of the spectrum. The highlighted area was used for the calculation of relative error at high energy in table II. From the fit to the dataset, it is clear that the error is larger for structures with thicker top film.

TABLE II. *Relative error in fitting the sub-surface signal at high energy for four structures. The interface is modeled as infinitely sharp in the simulations. The relative error at high energy is calculated from 1850 eV as the area under the absolute residual error divided by the corresponding area under the experimental LEIS spectra.*

| Sample | Relative error at high energy (%) |
|---|---|
| 1.7 nm Si on W | 4.0 |
| 4.3 nm Si on W | 6.8 |
| 6.0 nm Si on W | 14.8 |
| 1.6 nm Si on Mo | 7.1 |

We observe an increase in the residual error when comparing the signal coming from interfacial W for the three samples. In the experiments, the higher the depth of the interface, the broader the energy distribution of the ion beam that reaches such compositional change (straggling). The simulation of the LEIS sub-surface signal disregards the contribution of electronic straggling, therefore the residual error in the fit increases as a function of the depth of the interface.

For the structure 1.7 nm Si on W (figure 4a), the relative error in fitting the W signal at high energy is 4.0%. Note that the interface is modeled infinitely sharp in the simulation, therefore only part of the error is attributed to the effect of electronic straggling, while another part is caused by the finite interface width in the experiment.

Since we are interested in characterizing the interface width, if we reduce the thickness of the top film to the lowest possible value while still achieving a top film that fully covers the substrate, we are able to reduce the contribution of the error due to electronic straggling and therefore get a realistic model for the intrinsic part of the spectrum which is sensitive to the interface width.

### *B. Qualitative comparison of interfaces*

To investigate whether the method of comparing LEIS sub-surface signals with the corresponding simulations is sensitive to the interface width, we compare the relative error at high energy obtained from two different structures, Si on W and Si on Mo. We make use of two bilayer structures with a similarly thin top film, 1.7 nm for Si on W and 1.6 nm for Si on Mo. The residual error due to straggling is expected to be similar in the two structures.

For the structure Si-on-W, we obtained a relative error at high energy equal to 4%. When considering another structure, if we assume the surface roughness to be similar, any larger variation between the experiment and simulation can be attributed to a broader interface. The result is shown in figure 5.

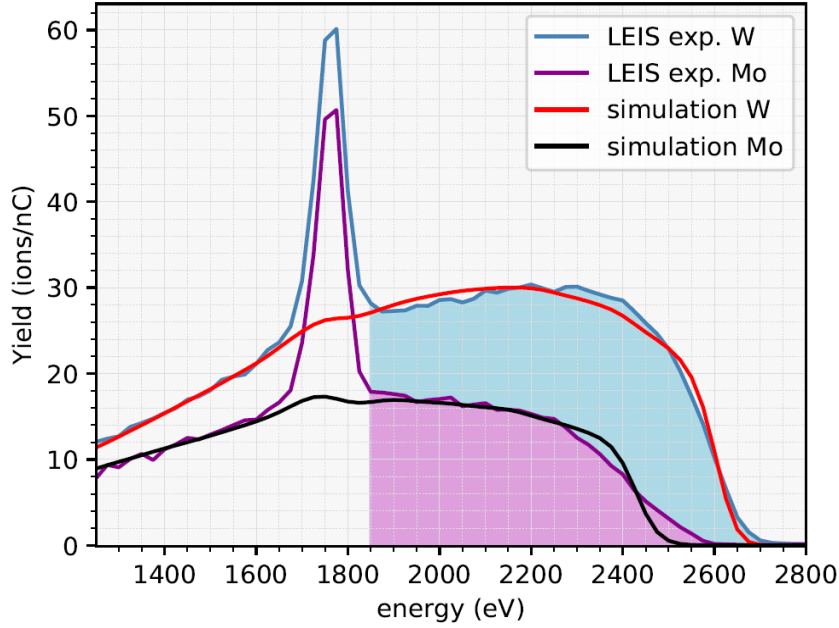

FIG. 5. LEIS experiment with 3keV He+ ions compared to the corresponding simulation for the structures 1.7nm Si on 20 nm W and 1.6 nm Si on 20 nm Mo. The interface is modeled as infinitely sharp in the simulation. The highlighted area was used to calculate the relative error at high energy in table II. It is clear from the presented data that the error for Si-on-Mo is larger than for Si-on-W, indicating a larger interface width for the Si-on-Mo system.

Comparing the interface signal for the two structures, the Si-on-Mo structure has a higher residual error, suggesting a broader interface. This is in agreement with what is predicted by previous studies on the two structures [10, 11] and by empirical rules based on atomic size difference, surface-energy difference, and mixing enthalpy developed by Chandrasekaran et al. [9]. From this qualitative analysis, the method of comparing LEIS sub-surface signals with the corresponding simulations appears sensitive to the interface width.

### *C. Measurement of interfaces*

To investigate whether it is possible to determine the width of a buried interface by comparing the experimental and simulated LEIS spectra, we focus on the structure Si-on-Mo which was investigated by the LEIS layer growth profile in the study [10].

We implement an interface layer in the TRBS simulations and study the relative error at high energy as a function of the thickness of the simulated interface. We increase the resolution of TRBS simulations by reducing the energy range to 1500-2700 eV. The corresponding energy resolution is a channel size of 6 eV.

As it is not known a priori what is the best model to describe the interface, we used two designs, a one-layer interface, and a four layers interface. From this, we test the sensitivity of the modeling to the variance in interface design. Figure 6 shows a sketch of how the interfaces are implemented in the simulations. The values of the parameters used for the simulations are listed in table III. Figure 7 shows the relative fitting error as a function of the total thickness of the simulated interface for the two cases. Figure 8 shows the simulated spectra corresponding to the best fit for the two cases.

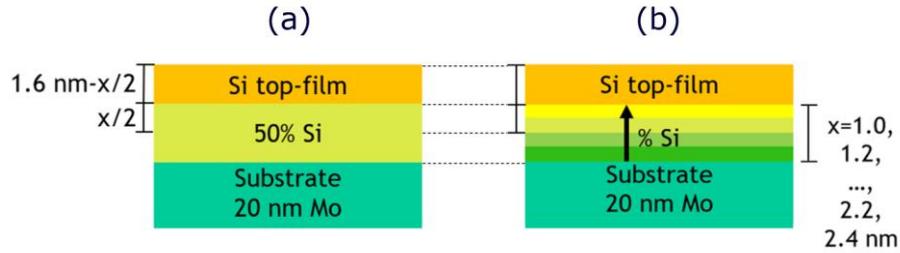

FIG. 6. Simulated layer stack, where the interface is modeled as a single layer (a), and as formed by four layers (b). The total thickness of the simulated interface, x, was modeled for discrete values between 1.0 to 2.4 nm. For an interface thickness x, Si thickness was reduced by a factor x/2 assuming the interface is allocated 50% inside the silicon film and 50% inside the Mo film. The simulation parameters used for each layer are reported in table III.

TABLE III. Simulated layer stack (from top to bottom) where the interface is modeled as a single layer and as formed by four layers. The composition, density, and ESC of the layers were defined through linear extrapolation between Si and Mo values.

| One-layer interface | | | |
| --- | --- | --- | --- |
| thickness (nm) | Composition (% of Si) | Density (g/cm3) | ESC |
| 1.6 nm – $x$/2 | 100 | 2.33 | 2.2 |
| $x$ | 50 | 6.31 | 2.1 |
| 20 nm | 0 | 10.28 | 1.9 |

| Four layers interface | | | |
| --- | --- | --- | --- |
| thickness (nm) | Composition (% of Si) | Density (g/cm3) | ESC |
| 1.6 nm – $x$/2 | 100 | 2.3 | 2.2 |
| $x$/4 | 80 | 3.9 | 2.1 |
| $x$/4 | 60 | 5.5 | 2.1 |
| $x$/4 | 40 | 7.1 | 2.0 |
| $x$/4 | 20 | 8.7 | 2.0 |
| 20 nm | 0 | 10.3 | 1.9 |

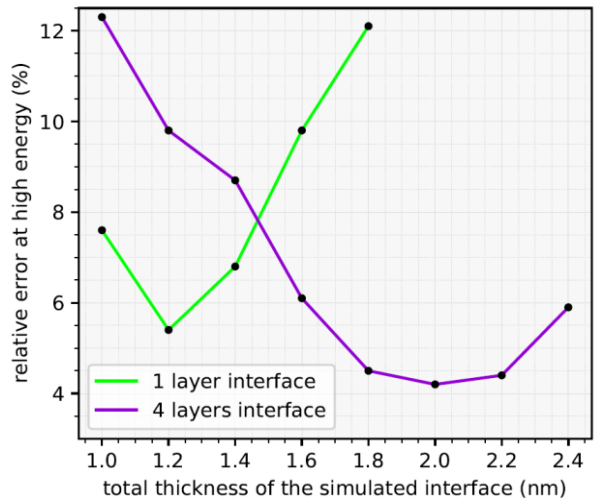

FIG. 7. Relative error at high energy as a function of the total thickness x of the simulated interface for 3 keV He ions on the structure 1.6 nm Si on 20 nm Mo. Two models were used for the interface as described in figure 6.

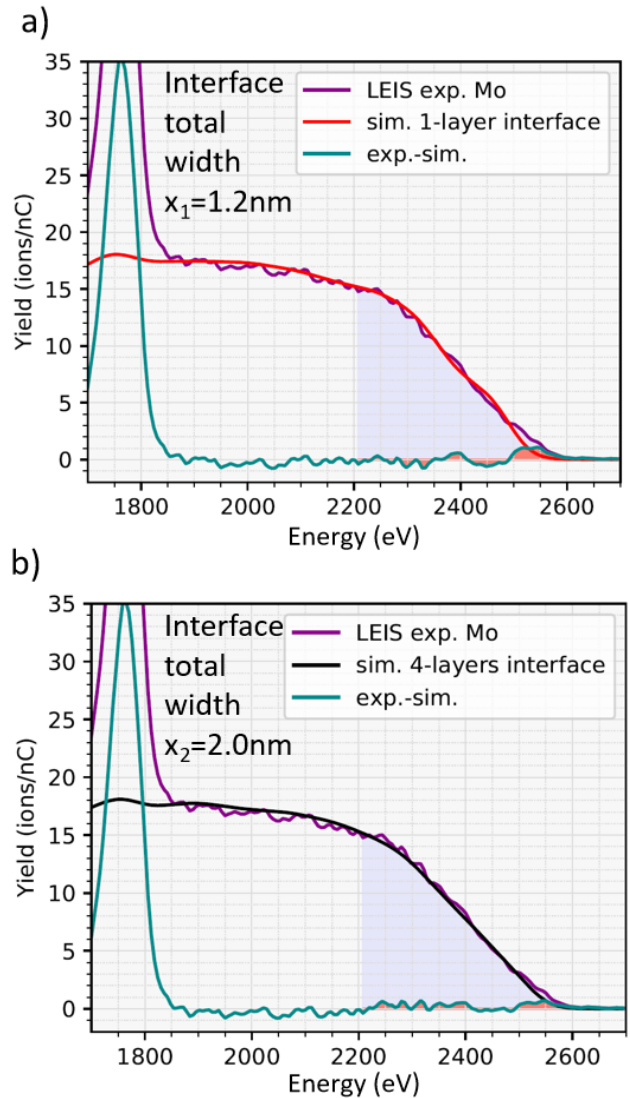

FIG. 8. LEIS experiment with 3 keV He ions compared to the corresponding simulation for the structure 1.6 nm Si on 20 nm Mo. a) the interface was modeled as one-layer containing Si and Mo as described in figure 6. b) the *interface was modeled as made by four layers as described in figure 6. The residual error is calculated as the difference between experiment and simulation for each point of the spectrum. The highlighted area corresponds to the area of deviation in figure 5 and was, therefore, used to calculate the relative error at high energy reported in figure 7.*

The four-layers model led to a minimum relative error of 4.2%. This is significantly smaller than the minimum relative error of 5.4% obtained by the one-layer model. Assuming a gradual compositional change in the structure, it is expected that the relative error decreases for an increasing number of layers in the model.

When the interface is modeled by four layers, the total thickness of the simulated interface yielding the best fit to the measured data is 2.0 nm (figure 7, purple). When the interface is modeled as one layer, the optimal value for the total thickness of the simulated interface is 1.2 nm (figure 7, green). Note that adding more steps in the simulated interface equals refining the fit towards a gradual compositional change which represents a realistic interface. Therefore, the total thickness of the simulated interface, x, increases as a function of the number of steps used for the simulation, as illustrated in the scheme in figure 9.

To retrieve the effective width σ (nm) of the two simulated interfaces, we fit them with an error function. With the one-layer model, we obtain σ=0.72 nm. With the four-layers model, we obtain σ=0.80 nm. Given the smaller relative error, the four-layers model is considered more accurate. However, it is important to notice that, even by modeling the interface with a single-layer, the difference in the final effective interface width is relatively small (0.08 nm).

Finally, the interface width measured with this method is likely to be an overestimate. This is due to the fact that straggling (acting like a smoothening factor) is not modeled by the uncorrected TRBS spectrum used for this study. For comparison, following the method described in [9], we extract the logistic function-like profile of the Si-on-Mo interface from the layer growth profile measured by Reinink et al. in the study [10]. The corresponding effective interface width is σ=0.79 nm.

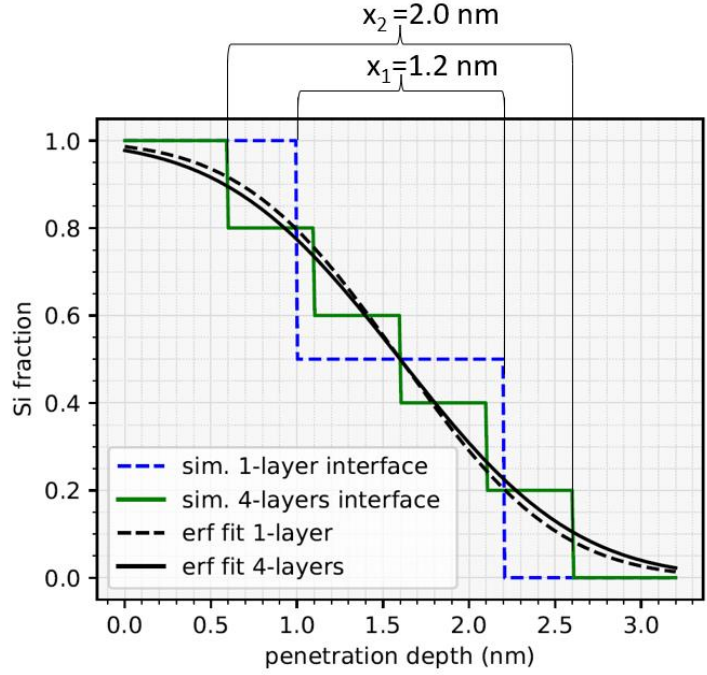

*FIG. 9. Model of an error function-like concentration profile with a one-layer interface and a four-layers interface. The total thickness of the interface resulting from the one-layer interface model, $x_1$, is smaller than the total thickness resulting from the four-layers interface model, $x_2$. When fitted with an error function, the two models lead to similar profiles.*

## D. Silicon sub-surface signal

The LEIS sub-surface signal (also called background) obtained with this method simulates the signal from all projectiles that are reionized at the surface after experiencing backscattering and charge transfer phenomena inside the target. The simulations in figures 3, 4, and 5 clearly shows a contribution of projectiles backscattered by silicon, the so-called silicon tail, in the energy range of silicon backscattering (below 1800 eV). In comparison, when the background of the LEIS spectrum is fitted with an error function or a polyline, there is no defined way to estimate the contribution of the silicon tail. This makes it difficult to establish a standard procedure to fit the background.

The background subtraction is a necessary step for the quantification of the area under surface peaks (surface quantification). We find that the simulation of the LEIS background with the described method might help in the process of establishing a standard procedure for background subtraction. However, a more detailed investigation is required for this purpose and that is beyond the scope of this paper.

## VII. CONCLUSIONS

The use of Low Energy Ion Scattering to <u>quantitatively</u> characterize buried interfaces was investigated. LEIS spectra contain depth-resolved information in the sub-surface signal. The latter can provide a relatively high yield when the structures are formed by heavy elements such as transition metals. In this study, we investigated structures of W/Si and Mo/Si thin films. The LEIS spectra provided qualitative information about the buried interfaces.

A methodology to assist the spectrum analysis with simulations has been explored. In the case of ultrathin films (<1.7nm) deposited on thick substrates, TRBS simulations can be used without a model for electronic straggling to simulate LEIS sub-surface signals. In the case of Si-on-W bi-layer structures, whose interface is expected to be relatively sharp, the relative error in fitting the sub-surface signal of the experimental spectra can be as low as 4%.

Excluding electronic straggling in the simulation leads to increasing residual error for increasing thickness of the top Si film. This result shows that models for electronic straggling should be depth dependent in the LEIS regime.

Simulations of LEIS sub-surface signals obtained by the presented method are sensitive to the interface width. For the structure Si-on-Mo, we obtained an optimal value for the interface width by introducing an interface layer of increasing width in the simulation. The resulting effective interface width of 0.8 nm ± 0.08 nm is in good agreement with the value of 0.79 nm, measured from the layer growth profile obtained by Reinink et al. [10].

This approach extends the use of LEIS to the characterization of buried interfaces without the need for sputter profiling. Interfaces play such an important role in the performance of thin films, that enabling a highly accurate and non-destructive measurement inside the structure is extremely valuable. Extending the study to other material systems is necessary to further assess the reliability and accuracy of the method.

## ACKNOWLEDGMENTS

This work has been carried out in the frame of the Industrial Partnership Program "X-tools," Project No. 741.018.301, funded by the Netherlands Organization for Scientific Research, ASML, Carl Zeiss SMT, and Malvern Panalytical. We acknowledge support of the Industrial Focus Group XUV Optics at the MESA+ Institute for Nanotechnology at the University of Twente.

## DATA AVAILABILITY

The data that support the findings of this study are available from the corresponding author upon reasonable request.

## AUTHOR DECLARATIONS

**Credits**

This article has been accepted by the Journal of Vacuum Science & Technology A. After it is published, it will be found at [Link](Link).

**Conflict of Interest**

The authors have no conflicts to disclose.

**Author Contributions**

**A. Valpreda**: Conceptualization (lead); Formal analysis (lead); Investigation (lead); Methodology (lead); Writing – original draft (lead); Writing – review & editing (lead). **A. Yakshin**: Conceptualization (supporting); Formal analysis (supporting); Investigation (supporting); Methodology (supporting); Writing – review & editing (supporting). **J. M. Sturm**: Conceptualization (supporting); Formal analysis (supporting); Investigation (supporting); Methodology (supporting); Writing – review & editing (supporting). **M. D. Ackermann**: Project administration (lead); Conceptualization (supporting); Formal analysis (supporting); Investigation (supporting); Methodology (supporting); Writing – review & editing (supporting).

## APPENDIX, Calculations of the maximum energy of Si sputtered atoms

We consider a system in which there is a collision cascade as formed by the following steps:

a. An incident projectile (1) of mass $m_1$ travels through the sample before the backscattering event. To estimate the case of maximum final energy we consider the minimum travel depth (and hence minimum stopping) of 3Å

b. the projectile (1) gets backscattered by a target atom (2) of mass $m_2$, which in this case is silicon. We assume that (2) is at rest before the collision.
c. The projectile (1) travels back after the backscattering
d. The projectile (1) kicks out a Si atom (3) at the surface

These steps can be described as follows.

The energy after the free path a. can be calculated as

$$E_a = E_0 - S\, d_a$$

(1)

Where $E_0$ is the primary energy of the ions, S is stopping calculated by SRIM software, $S[eV/Å]=4.9$, and $d_a$ is the distance travelled in the free path a., $d_a=3Å$.

The energy after the backscattering event b. can be calculated as

$$E_b = E_a \left( \frac{\cos\theta + [(m_2/m_1)^2 - \sin^2(\theta)]^{1/2}}{(1 + m_2/m_1)} \right)^2$$

(2)

Where θ is the angle between the incoming trajectory and the outgoing trajectory as defined in figure 10, $m_1$ is the mass of the projectile (He) and $m_2$ is the mass of the target Si atom.

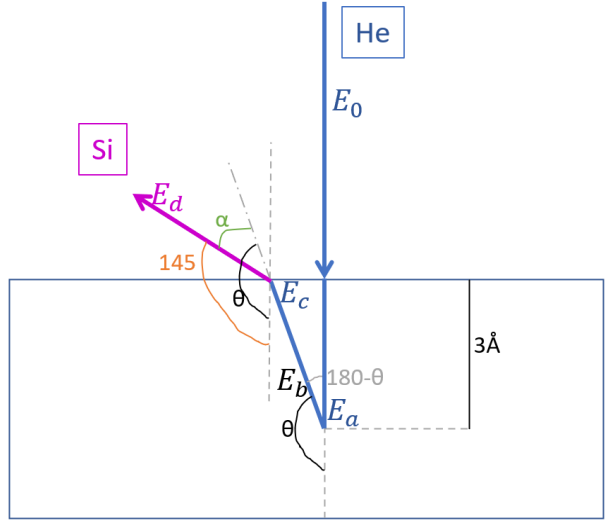

FIG. 10. Geometry of the collision cascade. We define the angle α as the angle between the trajectory of the backscattered He and the trajectory of Si after ejection. Note that the angle α can vary for a given final direction of the Si particle because multiple combinations of the backscattering angle θ and angle α can result in a Si

particle ejected in the direction of the detector. $E_0$, $E_a$, $E_b$, $E_c$ indicate the He kinetic energy at several stages in the collision cascade, while $E_d$ indicates the energy of the ejected Si atom.

The energy after the second free c. path can be calculated as

$$E_c = E_b - S \frac{d_a}{\cos(180° - \theta)}$$

(3)

The energy of the ejected Si atom after the sputtering event d. can be calculated as

$$E_d = E_c \left( \frac{4 m_1 m_2 \cos^2 \alpha}{(m_1 + m_2)^2} \right)$$

(4)

Equation 4 is described in [27] as equation 8.

We, therefore, get a formula for the energy of the Si sputtered atom as a function of α, the angle between the trajectory of the backscattered He and the trajectory of the Si atom after ejection. We plot the final energy $E_d$ as a function of α, for combinations of values of θ and α that lead to ejection of a Si atom in the direction of the analyser (i.e. at an angle of 145° with respect to the incoming He ion), and read the maximum value of $E_d$. Figure 11 shows the plot for the case of primary energy $E_a$=3keV. The corresponding maximum energy of sputtered silicon is $E_{max}$=771eV. This calculation was performed also for the primary energies $E_a$=4 keV, 5keV, and 6keV. The resulting maximum energies of sputtered silicon are reported in table VI.

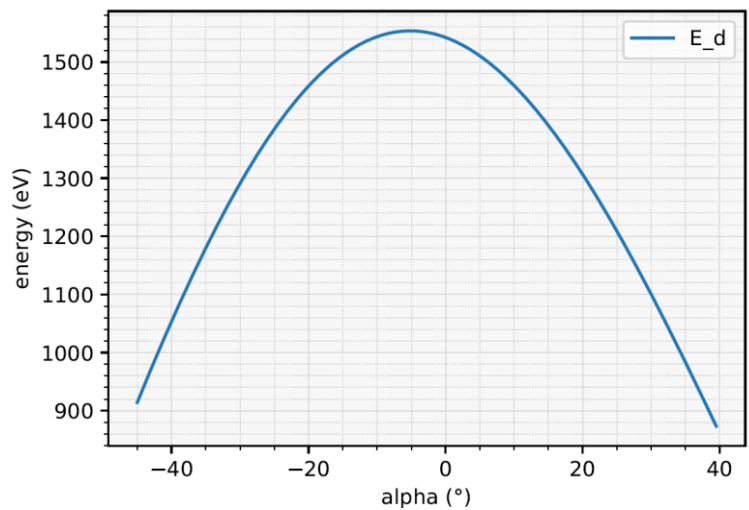

FIG. 11. Final energy $E_d$ of a Si particle ejected at a total angle of 145° with respect to the incoming He ion, as a function of the angle α in the case of primary energy $E_a$=3 keV.

*TABLE IV. Calculated value of the maximum energy for sputtered silicon atoms*

| Primary energy $E_0$ (eV) | Maximum energy of sputtered silicon (eV) |
|---|---|
| 3 | 771 |
| 4 | 1032 |
| 5 | 1293 |
| 6 | 1554 |